\documentclass[aps,prb,twocolumn,superscriptaddress]{revtex4}

\usepackage{graphicx}

\begin{document}


\title{ Enhanced coherent dynamics near a transition between neutral quantum-paraelectric and ionic ferroelectric phases in the quantum Blume-Emery-Griffiths model }


\author{ Kenji Yonemitsu }
\email[]{kxy@ims.ac.jp}
\affiliation{ Institute for Molecular Science, Okazaki 444-8585, Japan }
\affiliation{ Department of Functional Molecular Science, Graduate University for Advanced Studies, Okazaki 444-8585, Japan }

\date{\today}

\begin{abstract}
Nonequilibrium dynamics are studied near the quantum phase transition point in the one-dimensional quantum Blume-Emery-Griffiths model. Its pseudo-spin component $ S^z $ represents an electric polarization, and $ (S^z)^2 $ corresponds to ionicity, in mixed-stack charge-transfer complexes that exhibit a transition between neutral quantum-paraelectric and ionic ferroelectric (or antiferroelectric) phases. The time-dependent Schr\"odinger equation is solved for the exact many-body wave function in the quantum paraelectric phase. After impact force is introduced on a polarization locally in space and time, polarizations and ionicity coherently oscillate. The oscillation amplitudes are large near the quantum phase transition point. The energy supplied by the impact flows linearly into these oscillations, so that the nonequilibrium behavior is uncooperative. 
\end{abstract}

\pacs{78.20.Bh, 77.84.Jd, 77.80.-e, 64.60.Ht}
\keywords{metal-insulator interface, rectification}

\maketitle

\section{Introduction}

Correlated electron systems show various intriguing properties, which are not only scientifically interesting but potentially important for future applications. Their nonequilibrium dynamics have much room for searching novel functions. Among them, photoinduced phase transitions attract much attention because macroscopic properties are drastically changed by a cooperative effect often on an ultrafast time scale and in a controlled manner. \cite{nasu_book04,yonemitsu_jpsj06,yonemitsu_pr08} 

During photoinduced dynamics in quasi-one-dimensional electron systems, coherent oscillations are often observed, e.g., in the mixed-stack charge-transfer tetrathiafulvalene-$ p $-chloranil (TTF-CA) complex (in both short-time \cite{okamoto_prb04} and long-time \cite{iwai_prl02} behaviors), the quarter-filled-band charge-ordered organic (EDO-TTF)$_2$PF$_6$ (EDO-TTF=ethylenedioxy-tetrathiafulvalene) salt, \cite{chollet_s05} and potassium tetracyanoquinodimethane (K-TCNQ) crystals, \cite{okamoto_prl06} and their origins are theoretically investigated in itinerant electron models for TTF-CA (with short-time \cite{yonemitsu_jpsj04c} and long-time \cite{yonemitsu_prb06} behaviors),  (EDO-TTF)$_2$PF$_6$, \cite{onda_prl08} and K-TCNQ. \cite{maeshima_jpsj08} In most of the cases, electron-phonon interactions are substantial. In theoretical works, phonons are often treated as classical variables in order to reduce the numerical burden when the electronic degrees of freedom are taken into account. However, quantum nature is indispensable to maintaining coherent oscillations at not-too-low temperatures where the classical Maxwell distribution of thermal lattice fluctuations would considerably deteriorate the coherence. 

Here, we focus on the quantum phase transition realized in mixed-stack charge-transfer complexes composed of 4,4'-dimethyltetrathiafulvalene (DMTTF) and tetrahalo-$ p $-benzoquinones (QBr$_n$Cl$_{4-n}$). \cite{horiuchi_jacs01,horiuchi_s03} For $ n $=2, 3, and 4, quantum paraelectric behaviors are observed. The dielectric permittivity follows the so-called Barrett formula as a function of temperature, where the divergence is inhibited by quantum fluctuations and the permittivity is saturated at low temperatures. Recently, photoinduced reflectivity changes are measured. \cite{kimura_jps08a} The reflectivity in the energy range where it is sensitive to the ionicity change shows a large-amplitude oscillation near the quantum phase transition point. 

In this paper, nonequilibrium dynamics are calculated after impact force is introduced to a polarization in the one-dimensional quantum Blume-Emery-Griffiths model. Near the quantum phase transition point, the correlation length is long and quantum fluctuations are large because the ordering is marginally suppressed. In such a case, the impact induces large-amplitude oscillations in polarizations and ionicity. The cooperativity of the induced dynamics is measured by calculating the ratios of the energies of induced oscillations to the total-energy increment. Below, we will show that the dynamics near the quantum phase transition exhibit enhanced oscillations in amplitudes but their behaviors are uncooperative in the present model. 

\section{Model and Method \label{sec:model_method}}

Mixed-stack charge-transfer complexes have different types of degrees of freedom, i.e., ionicity and electric polarization, although they are strongly coupled. A global change in the ionicity does not alter the symmetry, but a finite average of electric polarizations breaks the inversion symmetry. They can be described by spin-one operators, where $ S^z_i $ takes either +1, 0, or $-$1. A neutral state at site $ i $ is represented by $ S^z_i $=0, while an ionic state with positive or negative polarization by $ S^z_i $=1 or $-$1. Here, a finite polarization is assumed at each ionic site. Because of the spin-Peierls instability, an ionic domain consisting of many sites would have dimerization-induced electric polarizations. However, a few molecules may not have a polarization even if they are ionic. Furthermore, an odd number of consecutive ionic (neutral) molecules in the neutral (ionic) background cannot be described by such a pseudo-spin state. In this sense, the present ``sites'' should be interpreted in a coarse-grained sense. 

Once such coarse graining is accepted, we can employ the Blume-Emery-Griffiths model, \cite{beg_pra71} but one should be aware of the following points. If we could integrate out electronic degrees of freedom to obtain some effective model, it might have some long-range interaction that originates from the spin-Peierls instability of one-dimensional half-filled correlated electron systems. We will not pursue nonequilibrium dynamics specific to particular materials, but discuss rather general aspects near quantum critical points. Then, we will limit ourselves to short-range interactions. We do not distinguish between electronic and phonon contributions to electric polarizations, so that the present ``sites'' are composites of them. 

The Blume-Emery-Griffiths model is successfully applied to TTF-CA. It explains the existence of a triple point, \cite{luty_epl02} which is experimentally observed. \cite{lemee_prl97} The inclusion of interchain Coulomb attraction and interchain elastic energy qualitatively explains the pressure-temperature phase diagram of TTF-CA. \cite{kishine_prb04} Relaxation processes in the neutral-ionic, paraelectric-ferroelectric phase transitions \cite{collet_s03,guerin_cp04} are discussed on the basis of its master equation. \cite{inoue_prb07}

Now, we extend this model to the mixed-stack charge-transfer complexes, DMTTF-QBr$_n$Cl$_{4-n}$, which show quantum phase transitions between neutral and ionic phases. \cite{horiuchi_jacs01,horiuchi_s03} For the spin-one model with short-range interactions, the way by which quantum natures are incorporated is not unique. Here, we assume that tunneling between ionic states with different polarizations is achieved always through a neutral state and represented by the operator $ S^x_i $ defined below. This quantum version is also successful in that the mean-field theory reproduces the dielectric permittivity in the neutral phase, \cite{yamashita_jpcs08} which is described by the Barrett formula. \cite{horiuchi_jacs01,horiuchi_s03} 

This quantum version of the Blume-Emery-Griffiths model is defined by 
\begin{eqnarray}
H_{\mathrm{QBEG}} & = &
-J \sum_i S^z_i S^z_{i+1}
-K \sum_i (S^z_i)^2 (S^z_{i+1})^2 \nonumber \\ & &
-P \sum_i (S^z_i)^2
-h \sum_i S^x_i
\;, \label{eq:model}
\end{eqnarray}
where $ S^z_i = ( \mid \! 1 \rangle \langle 1 \! \mid - \mid \! -1 \rangle \langle -1 \! \mid )_i $ and $ S^x_i = (\mid \! 1 \rangle \langle 0 \! \mid + \mid \! 0 \rangle \langle 1 \! \mid + \mid \! 0 \rangle \langle -1 \! \mid + \mid \! -1 \rangle \langle 0 \! \mid)_i/\sqrt{2} $ with $ \mid \! j \rangle_i $ being the eigenstate associated with $ S^z_i $=$ j $. The periodic boundary condition is imposed. The parameter $ J $ ($ K $) denotes the nearest-neighbor dipolar (quadrupolar) interaction, $ P $ the energy difference between neutral and ionic states, and $ h $ the neutral-ionic quantum tunneling amplitude. Under hydrostatic or chemical pressure, the crystal contracts and the Madelung energy stabilizes the ionic state relative to the neutral state. This would correspond to increasing $ P $ in the present model. In the $ P \rightarrow \infty $ limit, the states $ \mid \! 0 \rangle_i $ are completely suppressed, and the model is equivalent to the quantum Ising model (i.e., the spin-1/2 transverse Ising model).

We study dynamics in nonequilibrium conditions by introducing the perturbation to Eq.~(\ref{eq:model}) written as, 
\begin{equation}
H' = - \sum_i E_i(t) S^z_i
\;, \label{eq:impulse}
\end{equation}
where $ E_i(t) = E_{\mathrm{imp}} $ for $ i = 1 $ and $ 0 < t < t_{\mathrm{imp}} $ and $ E_i(t) = 0 $ otherwise. The time-dependent Schr\"odinger equation for the exact many-body wave function is numerically solved by expanding the exponential evolution operator with time slice $ dt $=0.01 to the 15th order \cite{yonemitsu_prb07,onda_prl08} and by checking the conservation of the norm and of the total energy for $ t > t_{\mathrm{imp}} $. 

The main purpose here is not necessarily to simulate photoinduced dynamics, but to study nonequilibrium dynamics near the quantum critical point from a general viewpoint, i.e., coherence and nonlinearity. By the above excitation, different polarization states at site $ i $ get different energies. It does not directly transit between different polarization states. Note that, in itinerant electron models, the gauge transformation relates photoexcitation through the modulation of site energies via a scalar potential to that through the modulation of transfer integrals via a vector potential. \cite{yonemitsu_jpsj05} More importantly, Eq.~(\ref{eq:impulse}) gives a spatially local excitation in contrast to realistic photoexcitation, which is regarded as almost a uniform excitation. Later, we will discuss how strongly coherence is maintained near the quantum phase transition point. The impact force introduced above is disadvantageous because the spatial coherence is initially deteriorated. Even in such a difficult condition, physical quantities will coherently oscillate near the quantum phase transition point. 

\section{Results \label{sec:result}}

\begin{figure}
\includegraphics[height=12cm]{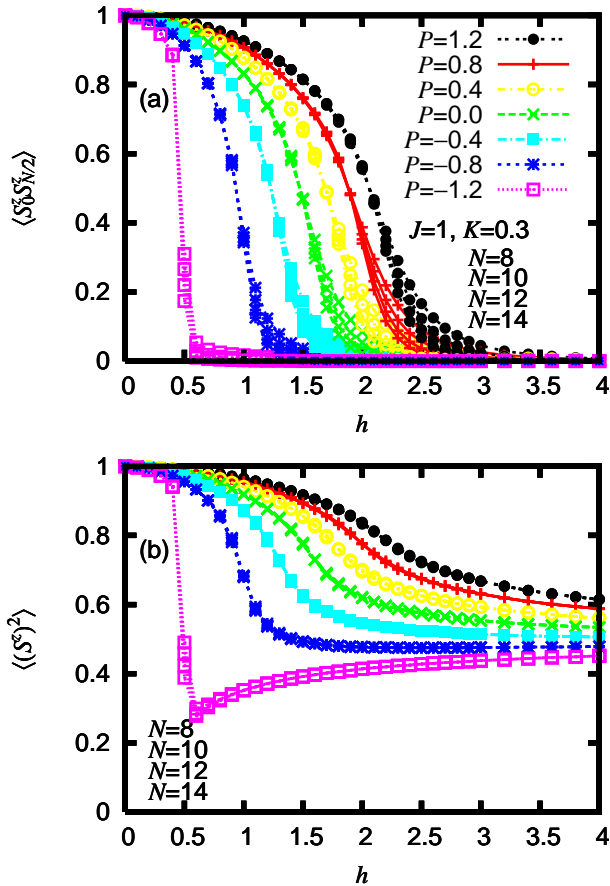}
\caption{(Color online) (a) $ S^z $-$ S^z $ correlation between farthest points and (b) expectation value of $ (S^z)^2 $, as a function of quantum tunneling amplitude $ h $, with $ J $=1, $ K $=0.3, and different $ P $ values. For each parameter set, the results of different system sizes, $ N $= 8, 10, 12, and 14, are shown. The curves with the steeper slope at the tail correspond to the results of the larger size. 
\label{fig:phase_diagram}}
\end{figure}

First of all, the ground-state properties are calculated and shown in Fig.~\ref{fig:phase_diagram}. 
The quantum tunneling between any states that are degenerate in the thermodynamic limit is inevitable in finite-size calculations and always leads to the vanishing expectation value of $ S^z $ even in the case where the symmetry is spontaneously broken in this limit. Then, we plot the $ S^z $-$ S^z $ correlation function between farthest points, i.e., $ \langle S^z_{N/2} S^z_{N} \rangle $ in Fig.~\ref{fig:phase_diagram}(a). It monotonically decreases with increasing $ h $ and monotonically increases with $ P $. Here, the results of different system sizes, $ N $= 8, 10, 12, and 14, are shown for each $ P $. The system-size dependence is visible for small $ \langle S^z_{N/2} S^z_{N} \rangle $. For large $ N $, it steeply decreases to zero. There would be a critical tunneling amplitude $ h_c $ above which $ \langle S^z_{N/2} S^z_{N} \rangle $ vanishes in the $ N \rightarrow \infty $ limit. The quantity $ h_c $ increases with $ P $. 

The ionicity $ \langle (S^z_i)^2 \rangle $ (at any $ i $) is shown in Fig.~\ref{fig:phase_diagram}(b). For $ P \geq -0.4 $, it monotonically decreases with increasing $ h $. For any $ P $, it becomes 0.5 in the $ h \rightarrow \infty $ limit because the ground state of $ -h \sum_i S^x_i $ is $ (1/2) \mid \! 1 \rangle_i +(1/\sqrt{2}) \mid \! 0 \rangle_i + (1/2) \mid \! -1 \rangle_i $. For $ P = -0.8 $, it decreases to a value below 0.5 (0.475) and then gradually increases to 0.5. With decreasing $ P $, the decrease of $ \langle (S^z_i)^2 \rangle $ with increasing $ h $ becomes steeper. For $ P = -1.2 $, it abruptly decreases between $ h $=0.5 and 0.6 and then increases to 0.5. In the thermodynamic limit, the transition from the ferroelectric phase ($ \langle S^z_i \rangle \neq 0 $) to the paraelectric phase ($ \langle S^z_i \rangle = 0 $) would be continuous at least for $ P \geq -0.8 $. Meanwhile, the transition seems discontinuous for $ P = -1.2 $. It indicates the presence of a tricritical point between $ -1.2 < P < -0.8 $ and $ 0.5 < h < 1 $. In any case, the dynamical behavior with increasing $ h $ above the transition point is similar for all $ P $, and the change with $ h $ is largest for $ P = -1.2 $. Then, we will show the time-dependent properties for $ P =-1.2 $ below. 

\begin{figure}
\includegraphics[height=12cm]{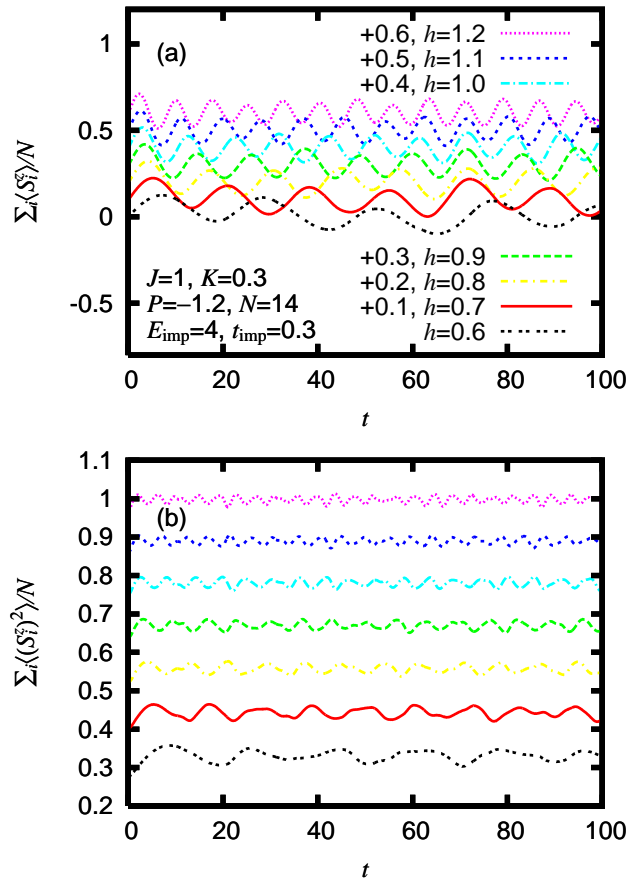}
\caption{(Color online) (a) Spatial average of $ S^z $ and (b) spatial average of $ (S^z)^2 $, as a function of time $ t $. The quantum tunneling amplitude $ h $ is varied from 0.6 with a step of 0.1, and the corresponding lines are shifted upward by 0.1. Other parameters are $ J $=1, $ K $=0.3, $ P $=$-$1.2, $ N $=14, $ E_{\mathrm{imp}} $=4, and $ t_{\mathrm{imp}} $=0.3. 
\label{fig:tunneling_dependence}}
\end{figure}

In the ground state, the polarization $ \langle S^z_i \rangle $ is absent at any $ i $ in the paraelectric phase. We have introduced impact force on  $ \langle S^z_i \rangle $ initially and at a particular site ($ i $=1), as described by Eq.~(\ref{eq:impulse}). Then, the polarization at site 1 grows at first. Owing to the coupling $ J $ and the periodic boundary condition, the polarizations grow successively at sites 2 and $ N $, at sites 3 and $ N-1 $, and so on. They oscillate with different phases and interfere with each other, so that their spatiotemporal pattern is complicated. Because the parameter set used here is in the paraelectric phase, their spatial average do not grow but oscillate about zero, as shown in Fig.~\ref{fig:tunneling_dependence}(a). 
Here, the lines with different $ h $ values are shifted vertically by different lengths. As the transition point is approached, the amplitude becomes large, and the period becomes long because the relevant excitation energy decreases. The dynamics of polarizations of course affect those of ionicity. The dynamics of ionicity are more complicated than those of polarizations because Eq.~(\ref{eq:impulse}) makes an impact directly on the polarization. Despite this, the spatial average of ionicity also oscillates, as shown in Fig.~\ref{fig:tunneling_dependence}(b), though its shape is not as sinusoidal as that of polarizations. There are similarities between the dynamics of ionicity and those of polarizations: as the transition point is approached, the oscillation amplitude becomes large, and the oscillation period becomes long. A detailed comparison will be made later. 

\begin{figure}
\includegraphics[height=12cm]{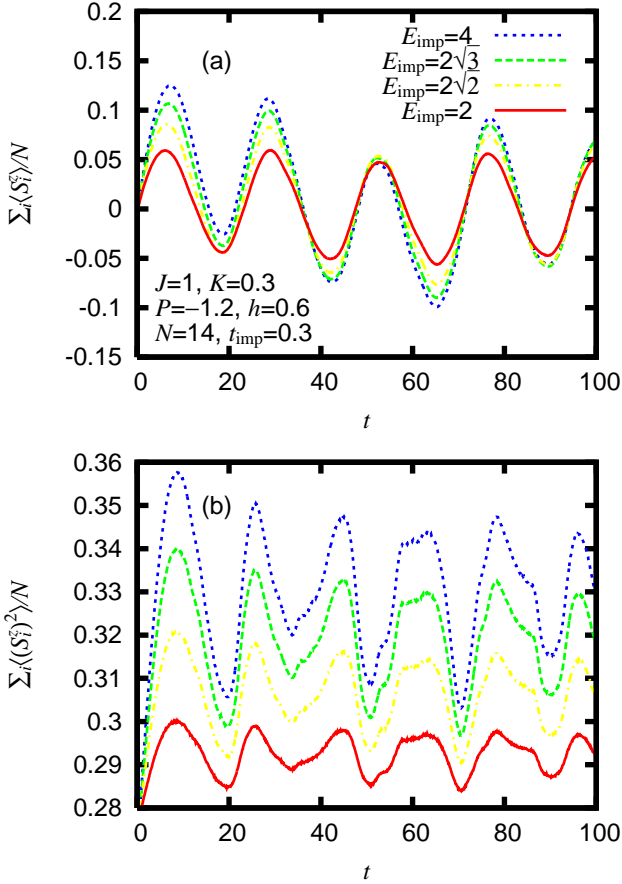}
\caption{(Color online) (a) Spatial average of $ S^z $ and (b) spatial average of $ (S^z)^2 $, as a function of time $ t $, with different $ E_{\mathrm{imp}} $ values. Other parameters are $ J $=1, $ K $=0.3, $ P $=$-$1.2, $ h $=0.6, $ N $=14, and $ t_{\mathrm{imp}} $=0.3. 
\label{fig:impulse_dependence}}
\end{figure}

In order to see nonlinearity or linearity of the nonequilibrium dynamics, its impact-strength dependence is important. Figure~\ref{fig:impulse_dependence} shows the $ E_{\mathrm{imp}} $-dependence of the time evolution of the spatially averaged quantities. 
Here, the lines with different $ E_{\mathrm{imp}} $ values are {\em not} shifted vertically. Because the increment in the total energy $ \Delta E = \langle H_{\mathrm{QBEG}} \rangle_{t=t_{\mathrm{imp}}} - \langle H_{\mathrm{QBEG}} \rangle_{t=0} $ is approximately proportional to $ E_{\mathrm{imp}}^2 $, we vary $ E_{\mathrm{imp}}^2 $ with a step of 4. With increasing $ E_{\mathrm{imp}} $, the oscillation amplitude becomes large, but the oscillation period does not change. The overall shape does not vary so much at least in this $ E_{\mathrm{imp}} $-range. The oscillation amplitude of the spatially averaged ionicity increases almost in proportion to $ E_{\mathrm{imp}}^2 $ [Fig.~\ref{fig:impulse_dependence}(b)], while that of the spatially averaged polarization increases sublinearly with $ E_{\mathrm{imp}}^2 $ [Fig.~\ref{fig:impulse_dependence}(a)]. 

\begin{figure}
\includegraphics[height=12cm]{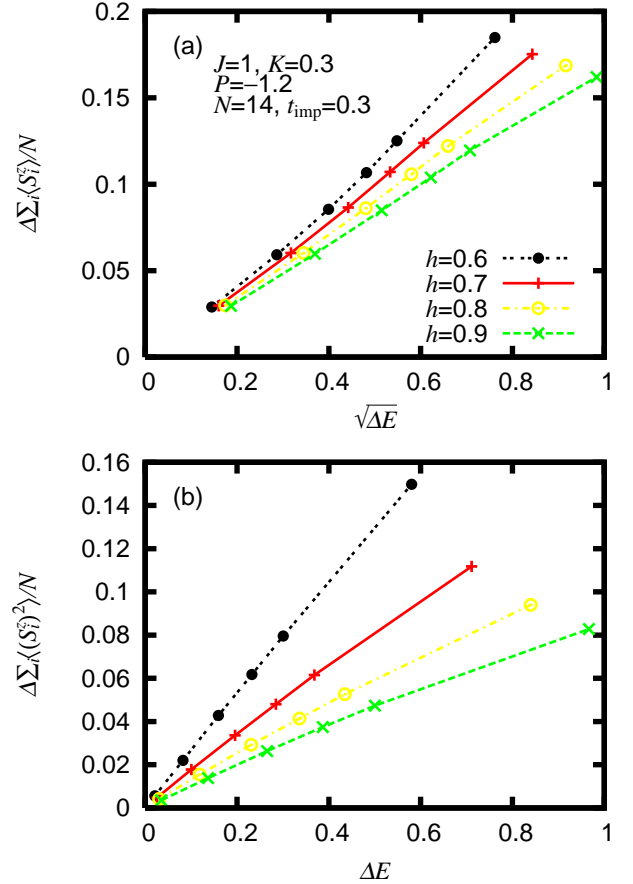}
\caption{(Color online) (a) Amplitude of oscillation in spatial average of $ S^z $, as a function of square root of energy supplied, and (b) amplitude of oscillation in spatial average of $ (S^z)^2 $, as a function of energy supplied. With decreasing tunneling amplitude $ h $ toward the quantum phase transition point, the slope becomes steep. Other parameters are $ J $=1, $ K $=0.3, $ P $=$-$1.2, $ N $=14, and $ t_{\mathrm{imp}} $=0.3. 
\label{fig:linearity}}
\end{figure}

The dependences of these oscillation amplitudes on the total-energy increment $ \Delta E $ are shown in Fig.~\ref{fig:linearity} for different $ h $ values. 
Here, the amplitudes are measured at the maxima on the first humps ($ t < 20 $). The oscillation amplitude of the spatially averaged polarization increases linearly with the square root of $ \Delta E $ [Fig.~\ref{fig:linearity}(a)], while that of the spatially averaged ionicity increases linearly with $ \Delta E $ [Fig.~\ref{fig:linearity}(b)]. In Fig.~\ref{fig:linearity}(a), the result for $ h = 0.6 $ visibly deviates from the linear function, but this is insignificant in terms of numerical accuracy. If the amplitude is measured at the maximum on the second hump instead, it becomes quite a linear function. It should be noted that the expectation value of Eq.~(\ref{eq:model}) contains a term that is quadratic with respect to polarizations, and terms that are linear and quadratic with respect to ionicity. No term is linear with respect to polarizations because of the inversion symmetry of the model. Therefore, the result shown in Fig.~\ref{fig:linearity} is a consequence of linear relations among the energy supplied by the impulse, the energy flow into polarizations, and the energy flow into increased ionicity. The nonequilibrium dynamics show uncooperative behaviors. The impact does not induce a transition into the ferroelectric phase. This linearity may be responsible for the observed coherence: the phase of the oscillating ionicity is maintained even if the impact is made on a polarization locally in space and time. If any cooperative effect were present, the ionicity would not oscillate about the initial value and its time evolution would show a more complicated behavior. 

What is interesting here is the enhancement of the oscillations near the quantum phase transition point: the linear coefficients are increased as the transition point is approached. The enhanced oscillation of the spatially averaged ionicity is conspicuous. In the quantum paraelectric phase (more generally, in quantum disordered phases), the interaction-induced ordering is suppressed by quantum fluctuations. Near the transition point, it is marginally suppressed. Then, any small perturbation would manifest this fact by showing a large response. There are large ferroelectric fluctuations accompanied by ionic fluctuations. The correlation length is long near the transition point. These facts enhance the coherent oscillations in polarizations and ionicity. These results are reminiscent of optical experiments on the perovskite-type quantum paraelectric SrTiO$_3$ \cite{takesada_jpsj03,hasegawa_jpsj03}, which show the photoinduced enhancement of the static dielectric permittivity near the quantum phase transition point. Although SrTiO$_3$ is a three-dimensional material, significant cooperative behaviors are not observed in the photoinduced changes. In the perovskite-type transition-metal oxides also, the electron-lattice \cite{ishihara_prb94a} and covalency \cite{ishihara_prb94b} contributions on the electronic polarizability are well recognized. 

\section{Summary \label{sec:summary}}

In order to approach the coherence in photoinduced dynamics of quasi-one-dimensional mixed-stack charge-transfer complexes near the transition between the neutral quantum-paraelectric phase and the ionic quantum-ferroelectric phase (within a chain), we theoretically study nonequilibrium dynamics of the one-dimensional quantum Blume-Emery-Griffiths model using exact many-body wave functions. The time evolution of the system is obtained by numerically solving the time-dependent Schr\"odinger equation after impact force is introduced on a polarization locally in space and time. Even though the impact is made locally, polarizations coherently oscillate about zero in the paraelectric phase, so that their oscillation survives after taking the spatial average. Furthermore, they induce a coherent oscillation in the spatially averaged ionicity. 

The induced oscillation is enhanced near the quantum phase transition point, so that the oscillation amplitude increases with decreasing tunneling amplitude toward the quantum phase transition point. The oscillation amplitude of the spatially averaged polarization increases linearly with the square root of the total-energy increment, while that of the spatially averaged ionicity increases linearly with the total-energy increment, implying linear relations among the energy supplied, the energy flow into polarizations, and the energy flow into increased ionicity. The linear coefficients are large near the quantum phase transition point because of large ferroelectric fluctuations accompanied by ionic fluctuations. Nevertheless the linearity implies uncooperative behaviors, so that the impact does not induce a transition into the quantum ferroelectric phase. This behavior is reminiscent of photoinduced behaviors of quantum paraelectric SrTiO$_3$, and it is also expected in the photoinduced dynamics in mixed-stack charge-transfer complexes near the quantum phase transition point. 

\begin{acknowledgments}
The author is grateful to H. Okamoto for sharing his data prior to publication and for enlightening discussions. 
This work was supported by Grants-in-Aid for Scientific Research (C) (No. 19540381), for Scientific Research (B) (No. 20340101), and ``Grand Challenges in Next-Generation Integrated Nanoscience" from the Ministry of Education, Culture, Sports, Science and Technology, Japan.
\end{acknowledgments}

\bibliography{qbeg08b_1}

\end{document}